\documentstyle[epsfig,psfig,aps,twocolumn]{revtex}

\newcommand{\rv}{{\bf r}}

\newcommand{\qv}{{\bf q}}
\newcommand{\kv}{{\bf k}}

\newcommand{\beq}{\begin{equation}}
\newcommand{\eeq}{\end{equation}}
\newcommand{\bea}{\begin{eqnarray}}
\newcommand{\eea}{\end{eqnarray}}

\renewcommand{\>}{\rangle}

\newcommand{\commentout}[1]{{}}

\begin{document}

\draft
\preprint{}
\title{Topological phase preparation in a pair of
atomic Bose-Einstein condensates}
\author{J. Ruostekoski}
\address{Abteilung f\"ur Quantenphysik,
Universit\"at Ulm, D-89069 Ulm, Germany}
\date{\today}
\maketitle
\begin{abstract}
We propose a method of generating skyrmion vortices
in a pair of Bose-Einstein condensates occupying two internal
states of the same atom.
We show that a variety of different periodic arrays of vortices 
may be prepared 
with an appropriate superposition of orthogonal standing 
electromagnetic waves inducing a coherent coupling between 
the two condensates.

\end{abstract} \pacs{03.75.Fi,05.30.Jp}

Since the first observations of Bose-Einstein condensation in dilute
atomic gases \cite{AND95,BRA95,DAV95} one important goal has been
to develop a technology to investigate
vortices in atomic Bose-Einstein condensates (BECs). The quantized
vortices in atomic BECs are analogous to the quantized circulation 
in superfluid liquid helium \cite{LIF80}. In this paper we propose 
a method of generating different periodic patterns 
of skyrmion vortices and solitary waves in a pair of atomic BECs. 
The two condensates
are coupled by electromagnetic (em) transitions. The em fields form
a standing wave configuration with topological singularities at
zero-field points. The em field amplitude couples to the relative
phase between the two BECs through the atomic transitions.
As a result the Rabi oscillations 
of atoms between the two internal states generate an array of 
topological phases in the two condensate system.

A BEC is expected to exhibit vorticity in a rotating harmonic trap if
the trap is anisotropic in the plane of rotation
\cite{DAL96,FET98,BUT99,STR99,FED99}.
It has also been proposed that stirring a BEC in a nonrotating 
configuration by a laser beam \cite{JAC98,CAR99,RAM99} or by an 
optically-induced potential \cite{MAR98} may generate vortices.
Topological defects may also emerge as a result of a rapid
condensation \cite{ANG99} or in self-interference measurements 
\cite{RUO99e}. 
Recently, it has been suggested that 
vorticity could be imprinted by imaging a BEC
through an absorption plate \cite{DOB99,HEL99}. Phase-imprinted dark
solitons have been observed in an atomic BEC \cite{HEL99,BUR99}.

While the previously explained schemes typically involve a BEC in a
single internal atomic state
several papers \cite{BOL98,MAR97,DUM98} have also considered the 
possibility of transferring an atomic population from a nonrotating 
ground state to a vortex state by means of a Laguerre-Gaussian laser 
beam consisting of photons with a nonvanishing orbital angular momentum.
In a very recent work Williams and Holland \cite{WIL99} have proposed
a scheme where two internal states are coupled and the trapping potentials
of these states are mechanically rotated. This technique has been
experimentally realized to demonstrate the formation of a vortex in 
a two-component BEC \cite{MAT99}. 

In this paper we propose a method of generating vortices which is 
related to the previously described schemes \cite{BOL98,MAR97,DUM98,WIL99} 
involving internal atomic transitions. 
We consider two internal states of the same atom
which are coupled by em fields via one or multiphoton transitions.
The em fields consist of two or more orthogonal standing waves. With an
appropriate wave configuration we obtain a spatially-dependent
transition strength with angular momentum singularities at zero-field
points. As a result of the spatially-dependent driving the internal
and external dynamics of the atoms are coupled: Around every zero-field
point we generate for the two-component BEC a topological phase with 
an integer multiple of $2\pi$ phase winding. The proposed scheme has
several advantages: It may be possible to prepare high-quality vortices
with little noise. The location of an individual vortex may be 
accurately controlled by shifting the zero-field point. 
We may also engineer a large selection of different periodic arrays of 
vortices in the two-condensate system, where the periodicity is 
determined by the driving em fields.
Moreover, with moving standing em waves we
can move the positions of the zero-field points, and therefore, also
the vortex array in a controlled way.

The dynamics of the two-component BEC occupying internal levels
$|1\>$ and $|2\>$ follows from the coupled
Gross-Pitaevskii equation (GPE) 
\begin{mathletters}
\bea
i \hbar \dot\psi_1 &=& \left( H_1^0
+\kappa_{11} |\psi_1|^2 +\kappa_{12} |\psi_2|^2\right)
\psi_1+\hbar\Omega^* \psi_2\,,\\
i \hbar \dot\psi_2 &=& \left(H_2^0
+\delta+\kappa_{22} |\psi_2|^2 +\kappa_{12} |\psi_1|^2\right)
\psi_2+\hbar\Omega \psi_1\,.
\eea
\label{gpe}
\end{mathletters}
Here the kinetic energy and the trapping potential are introduced
in $H_i^0$:
\beq
H_i^0\equiv -{\hbar^2\over 2m}{\bbox \nabla}^2+{1\over2} m\omega_i^2 
(x^2+\alpha_i^2 y^2+\beta_i^2 z^2)\,.
\eeq
We have also defined the coefficients of the nonlinearities as
$\kappa_{ij}\equiv 4\pi\hbar^2 a_{ij}N/m$. Here $a_{ii}$ denotes
the intraspecies scattering length in internal level $|i\>$ and
$a_{12}$ stands for the interspecies scattering length. 
The Rabi frequency, $\Omega(\rv)$, describes the strength of the
coupling between the two internal levels. The total
number of BEC atoms, the atomic mass, and the detuning of the em
fields from the resonance are denoted by $N$, $m$, and $\delta$,
respectively.

We assume that the Rabi frequency, $\Omega(\rv)$, 
is formed by two appropriately phase-shifted 
and orthogonal standing em waves, and has the following form:
\beq
\Omega(\rv)={\Omega_0\over\sqrt{2}} 
\{ \sin[k(x-\bar{x})]-i\sin[k(y-\bar{y})]\} \,,
\label{ome}
\eeq
where $k$ denotes the wave number of the em field. When the phase
shifts are equal to zero, $\bar{x}=\bar{y}=0$, the coupling vanishes
at $x_n=y_n=n\pi/k$, for $n=0,\pm1,\pm2,\cdots$. 
In the close neighborhood of the vanishing Rabi frequency,
$|k x_n-n\pi|\ll1$ and $|k y_m-m\pi|\ll1$, we obtain
\beq
\Omega(\rv)\simeq (-1)^{n}{\Omega_0 k\rho\over\sqrt{2}} e^{-\xi i\phi}\,,
\label{topph}
\eeq
with $\xi\equiv (-1)^{n+m}$.
Here $\rho\equiv\sqrt{x^2+y^2}$ is the radial coordinate in the $xy$ 
plane and $\phi$ is the corresponding polar angle. In Eq.~{(\ref{gpe})} 
the phase of the
Rabi frequency couples to the relative phase between the two condensates.
The phase of $\Omega(\rv)$ close to a zero-field point in Eq.~{(\ref{topph})}
has the form of the quantized circulation with the unit winding number.
We show that the coupling between the em and matter fields establishes
a topological relative phase between the two BECs.

To demonstrate the formation of vortices we numerically integrate GPE 
in two spatial dimensions. 
For simplicity, we assume that the traps are isotropic ($\alpha_i=1$),
and that the trapping frequencies and the potential minima of the two 
internal states are equal.
As an initial state we assume a nonrotating
ground state for a BEC in level $|1\>$ and an unoccupied level $|2\>$.
In the limit of strong self-interaction energy the initial
state of $\psi_1$ may be approximated by the Thomas-Fermi solution in 2D:
$\psi_1(\rho)=[2(R^2-\rho^2)/(\pi R^4)]^{1/2}$, for $R\geq\rho$, and
zero otherwise. Here
$R\equiv l [4\kappa^{2D}_{11}/(\pi \hbar\omega l^2)]^{1/4}$ denotes the 2D
Thomas-Fermi radius of the BEC and $l\equiv [\hbar/(m\omega)]^{1/2}$
is the harmonic trap length scale.

\begin{figure}
\begin{center}
\leavevmode
\begin{minipage}{4.2cm}
\epsfig{
width=4.2cm,file=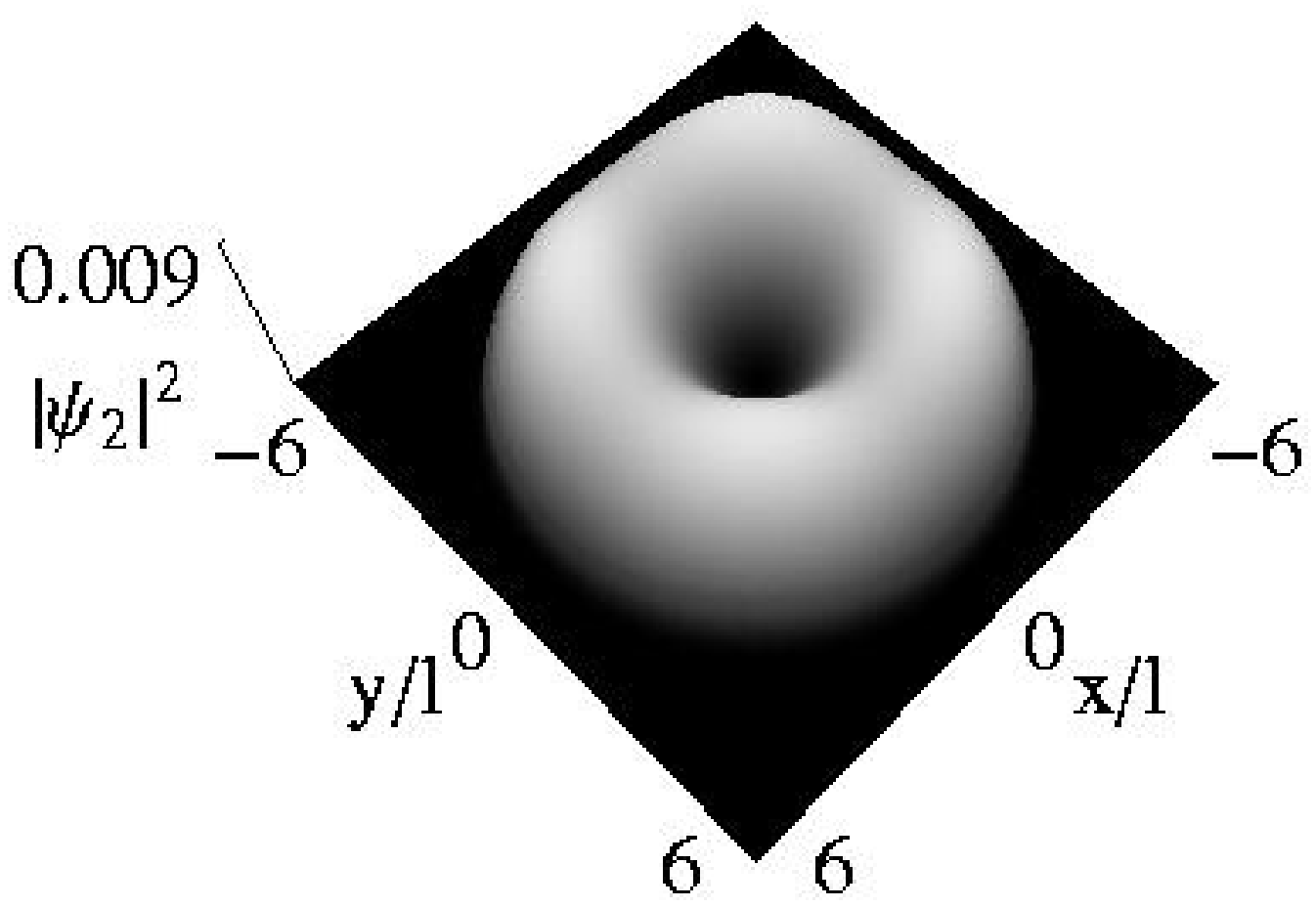}
\end{minipage}
\begin{minipage}{4.2cm}
\epsfig{
width=4.2cm,file=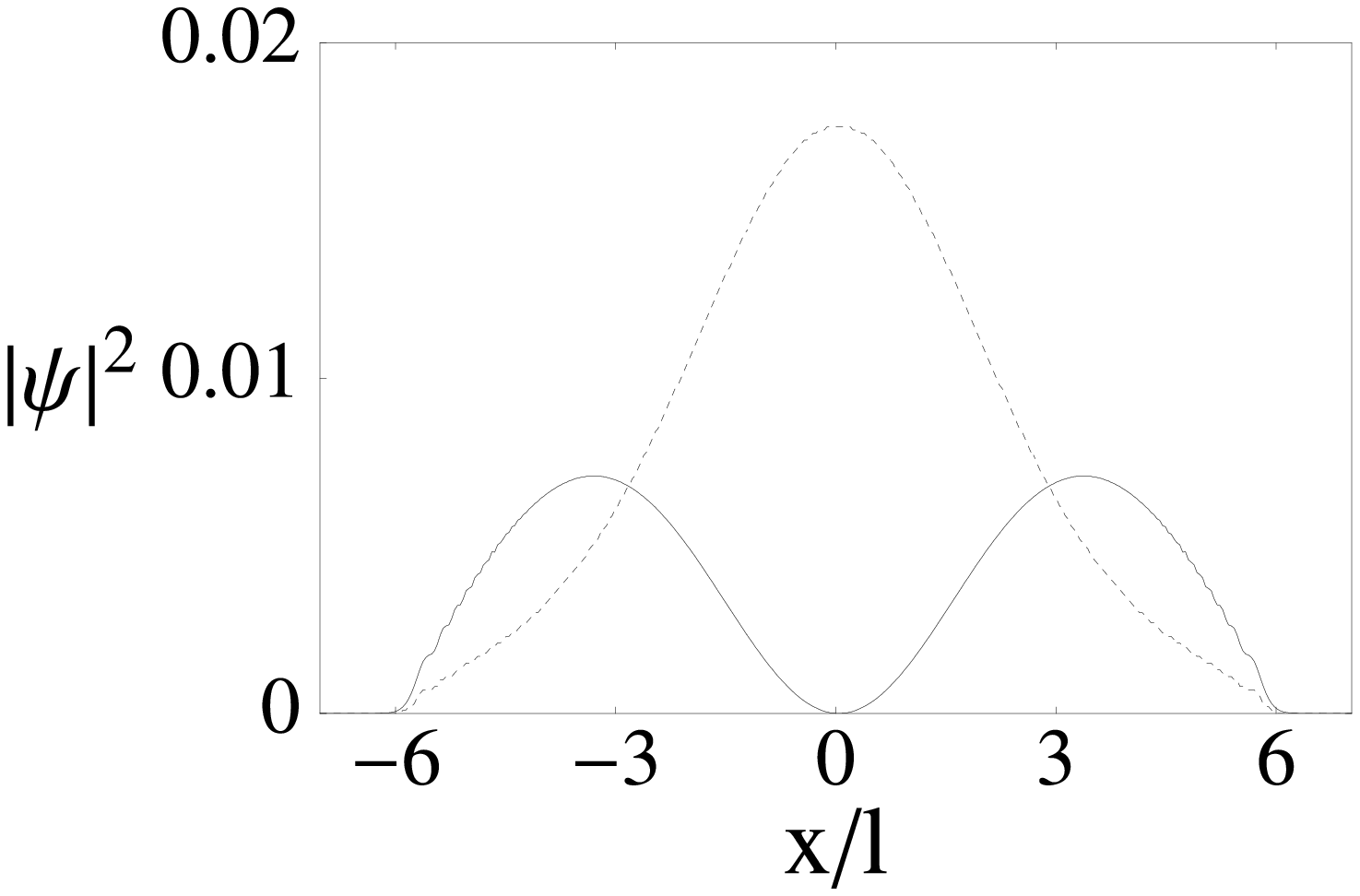}
\end{minipage}
\begin{minipage}{4.2cm}
\epsfig{
width=4.2cm,file=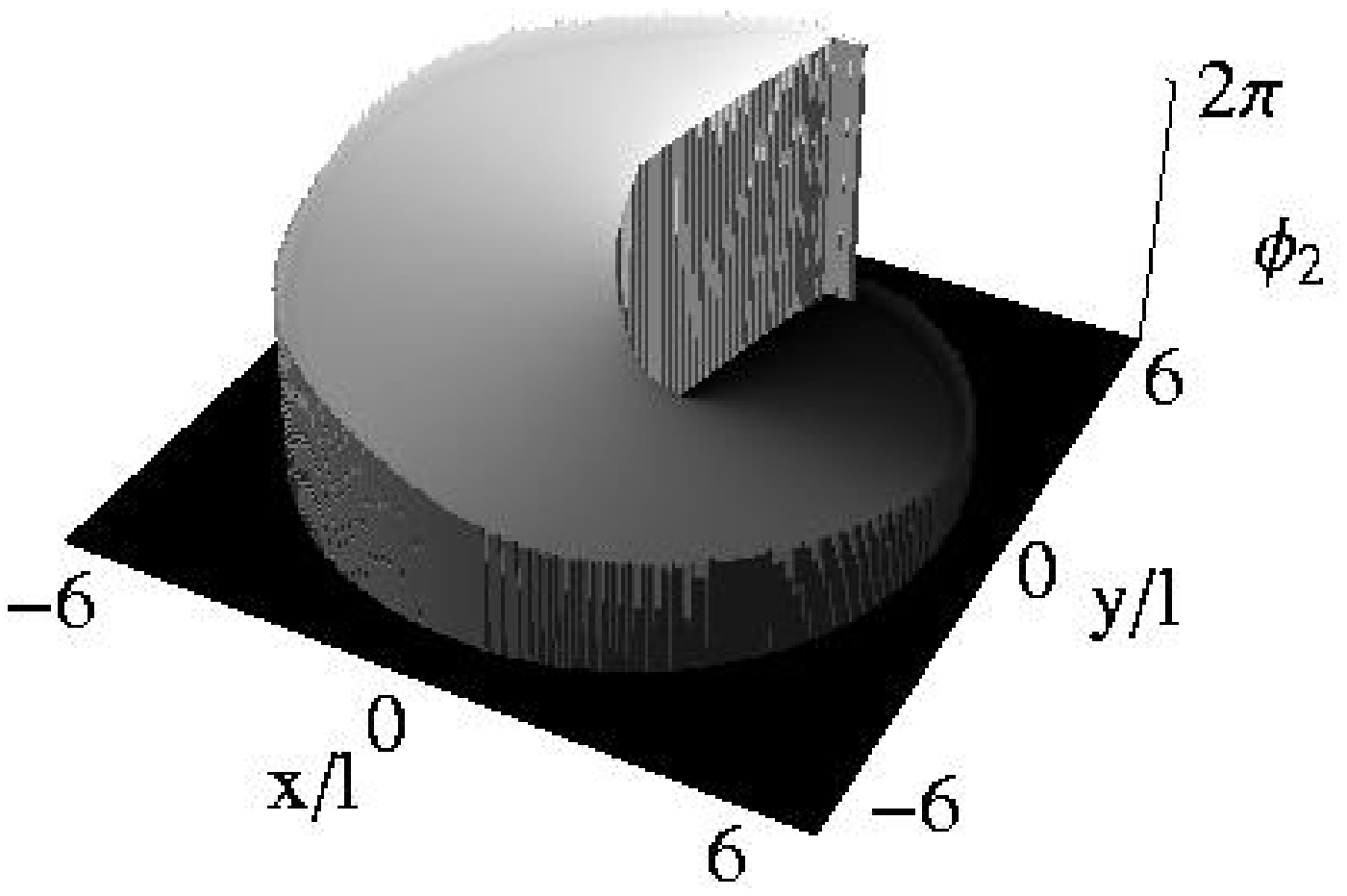}
\end{minipage}
\begin{minipage}{4.2cm}
\epsfig{
width=4.2cm,file=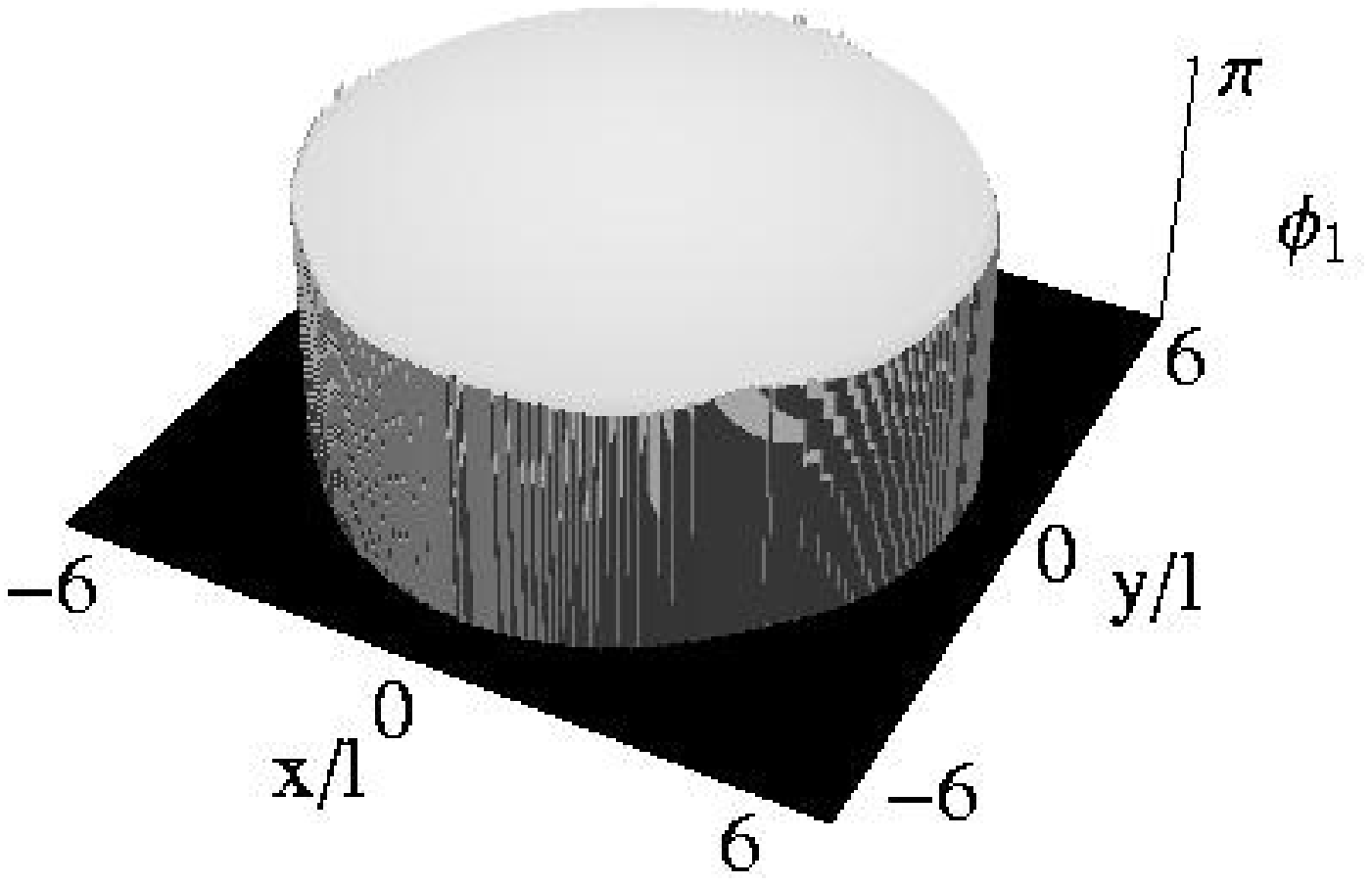}
\end{minipage}
\end{center}
\caption{
The preparation of a single vortex with the unit circular quantization. 
Before the driving em fields are turned on, a nonrotational
BEC occupies level $|1\>$. The em fields generate a vortex 
in level $|2\>$.
We show the density profile $|\psi_2(x,y)|^2$ (upper left) of atoms in
level $|2\>$, the density along the $x$ axis (upper right) in
levels $|2\>$ (solid line) and $|1\>$ (dashed line), and the 
phase profiles of levels $|2\>$, $\phi_2$, (lower left) and 
$|1\>$, $\phi_1$, (lower right) at a later time. The density
and phase of level $|2\>$ remarkably clearly display the 
characteristic properties
of a unit-quantized vortex with the $2\pi$ phase winding around the 
vortex core and the vanishing atom density in the center of the core. 
There is no vorticity in level $|1\>$.
}
\label{fig1}
\end{figure}

We choose the same ratio between the three
scattering lengths, 
$\kappa^{2D}_{11}:\kappa^{2D}_{12}:\kappa^{2D}_{22}::1.03:1:0.97$,
as for $^{87}$Rb states $|$F=1, m=-1$\rangle$ and $|$F=2, m=1$\rangle$
\cite{MAT99}. 
In the numerical calculations we use the nonlinearity 
$\kappa^{2D}_{12}/(\hbar\omega l^2)=1000$, the Rabi 
amplitude $\Omega_0=50\sqrt{2}\omega$, and the detuning $\delta=0$.
As a first example we create a single vortex with the unit topological
charge or the unit quantized
circulation. Here the value of the wavelength for the em fields
$\lambda=20l$ and $\bar{x}=\bar{y}=0$ in Eq.~{(\ref{ome})}, so that
only one field node is in the condensate. The atom population is initially
in level $|1\>$. The em fields start inducing transitions between the
two internal levels at time $t=0$. We decouple the two levels by 
turning off the em fields at time $t=0.02/\omega\simeq1.4/\Omega_0$. 
In Fig.~\ref{fig1} we display the density
$|\psi_2(x,y)|^2$ and the phase $\phi_2$ profiles of the BEC atoms in
level $|2\>$ at $t=0.0202/\omega$. We observe a vortex in level $|2\>$
with a vanishing atom density in the center of the vortex core and
the $2\pi$ phase winding around the vortex line demonstrating
the high quality of the preparation process. The phase profile $\phi_1$ of
$\psi_1$ is flat corresponding to a nonrotating state.

The vortex core in Fig.~\ref{fig1} is very large due to 
the atomic population 
$|\psi_1(x,y)|^2$ in level $|1\>$ which occupies the interior of the 
vortex and generates a mean-field repulsion as also displayed 
in Fig.~\ref{fig1}.
Therefore the presence of the vortex could possibly be directly
verified by imaging the density profile of the BEC in level $|2\>$
as in Ref.~\cite{MAT99}. 
The relative phase between two BECs has also a dramatic 
effect on the dynamical structure factor of the two-component 
system~\cite{RUO97b,PIT99} which may be observed, e.g., via the Bragg 
spectroscopy~\cite{STA99,BOL98b}.

\begin{figure}
\begin{center}
\leavevmode
\begin{minipage}{4.2cm}
\epsfig{
width=4.2cm,file=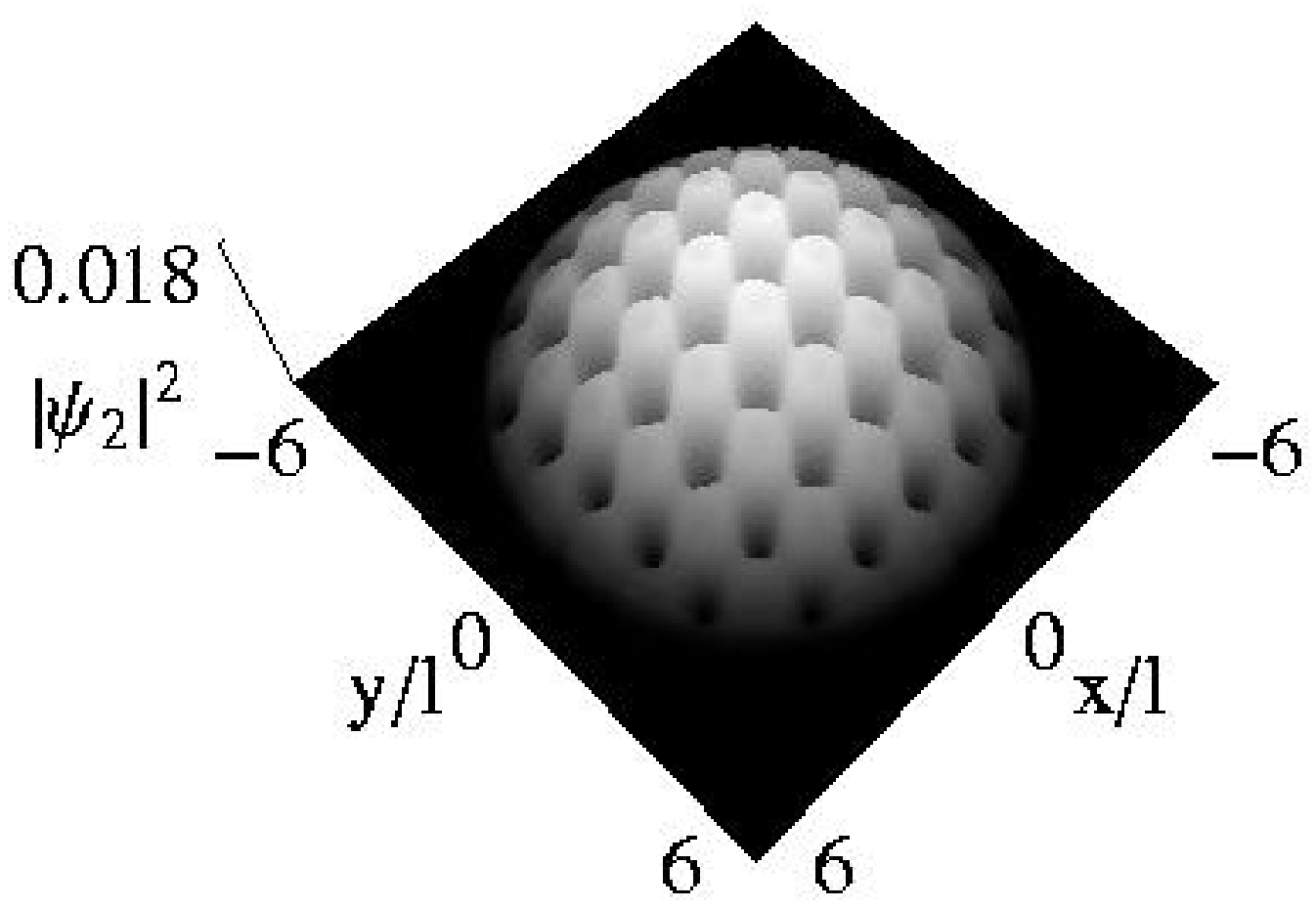}
\end{minipage}
\begin{minipage}{4.2cm}
\epsfig{
width=4.2cm,file=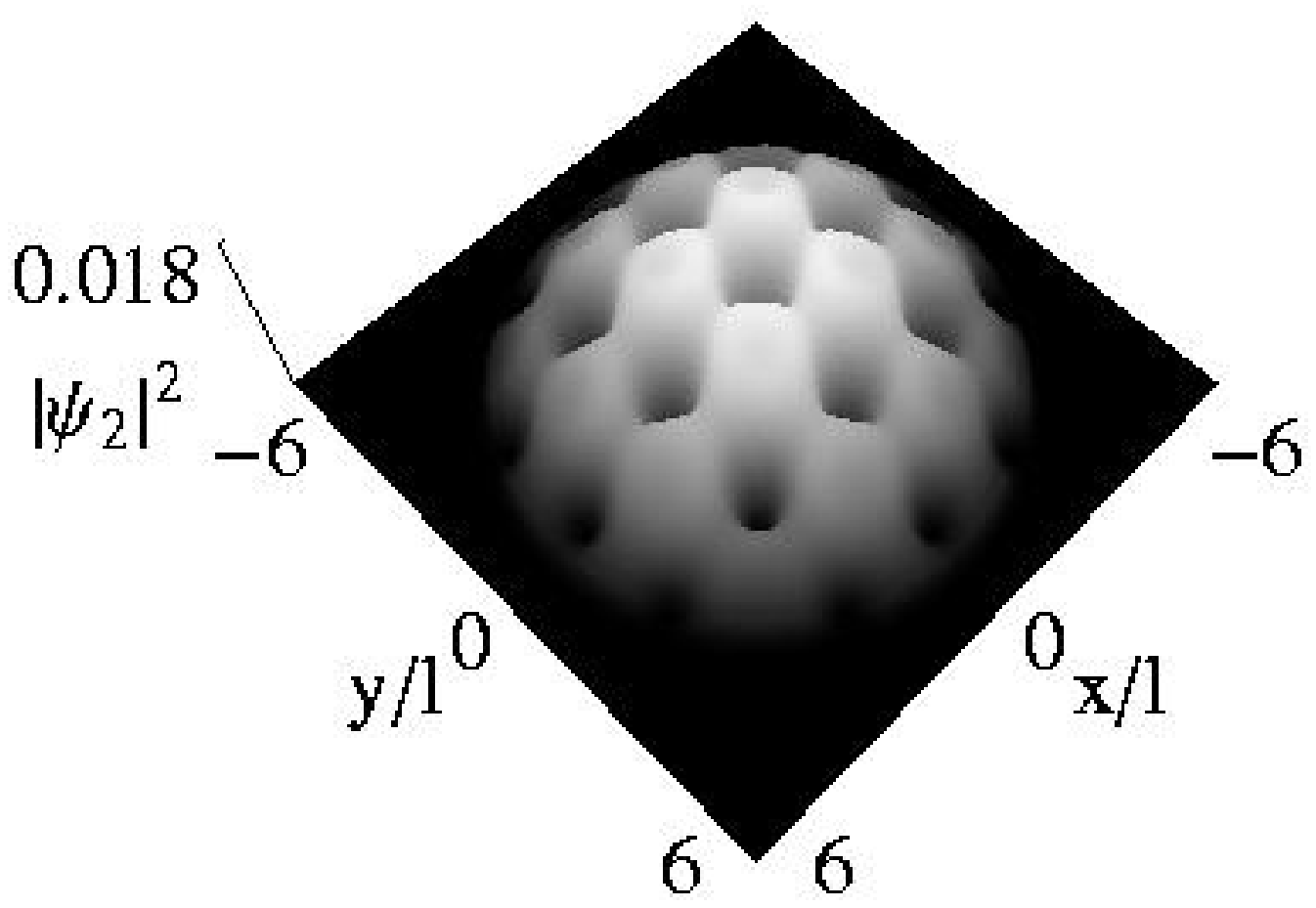}
\end{minipage}
\begin{minipage}{4.2cm}
\epsfig{
width=4.2cm,file=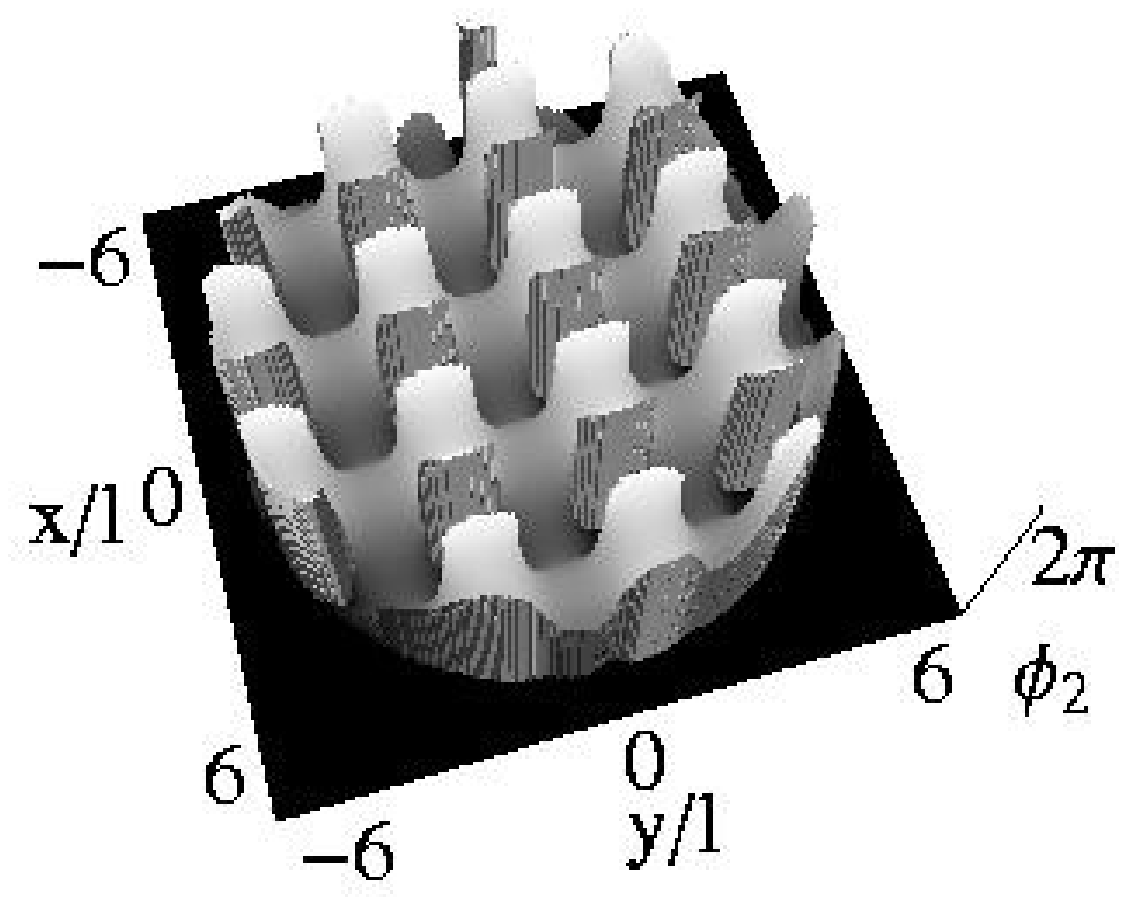}
\end{minipage}
\begin{minipage}{4.2cm}
\epsfig{
width=4.2cm,file=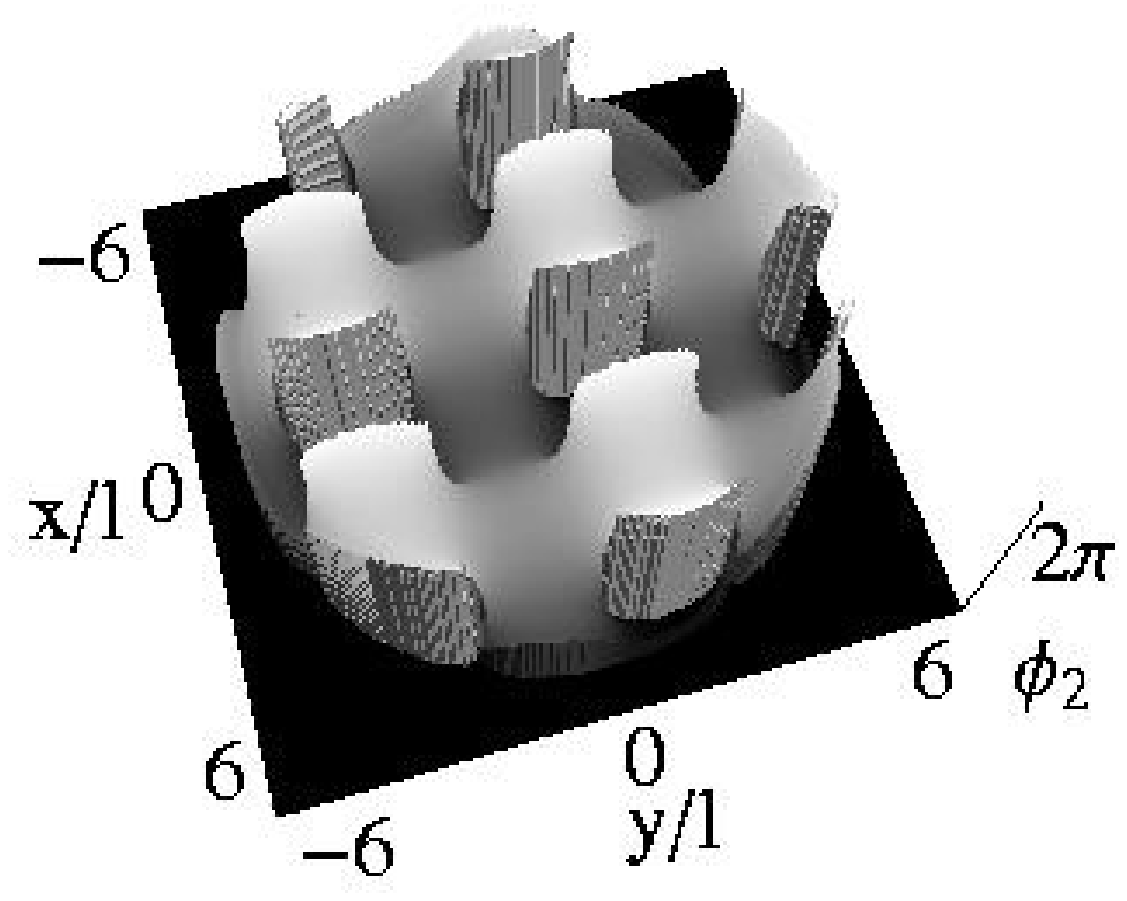}
\end{minipage}
\begin{minipage}{4.2cm}
\epsfig{
width=4.2cm,file=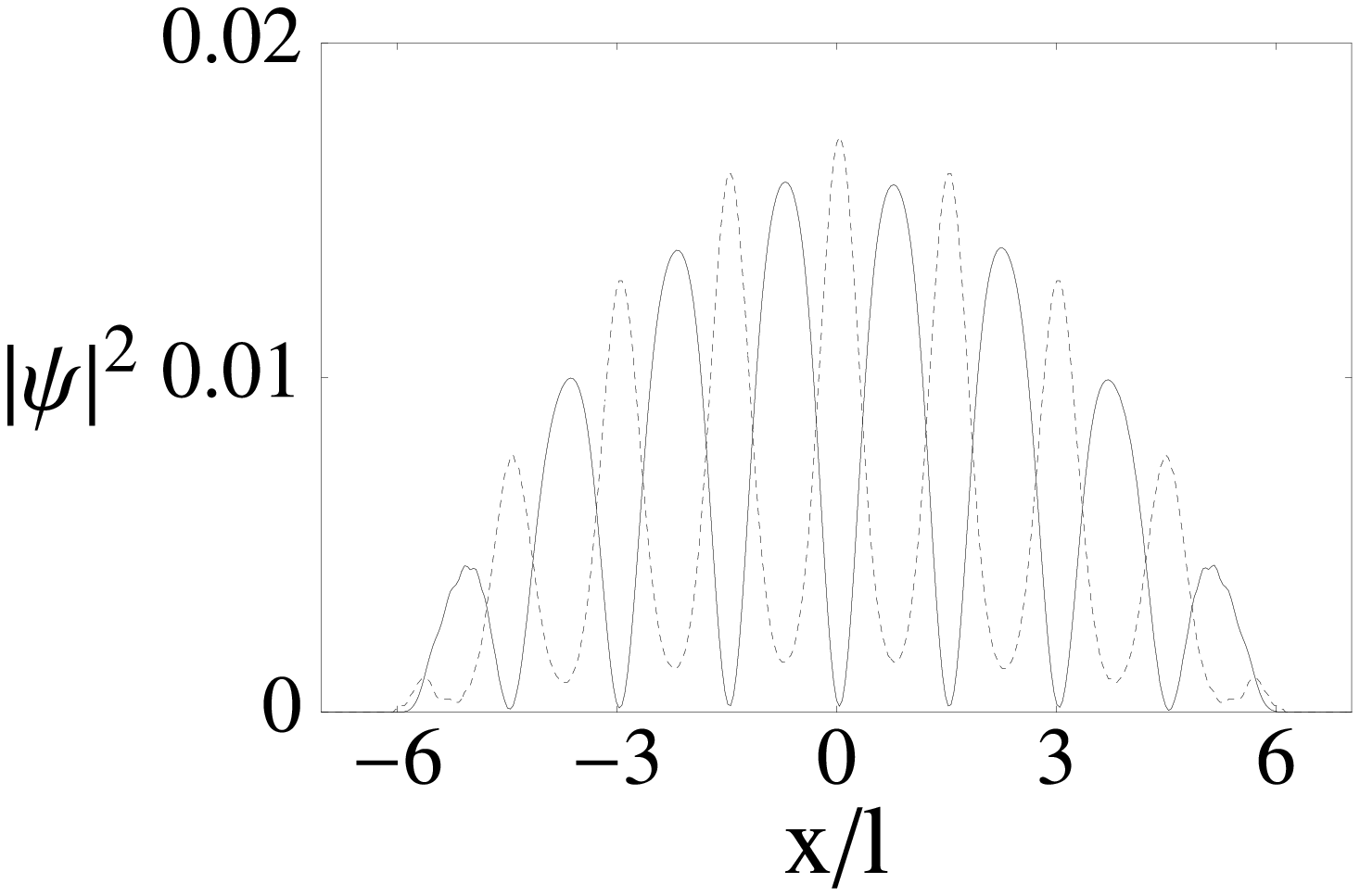}
\end{minipage}
\begin{minipage}{4.2cm}
\epsfig{
width=4.2cm,file=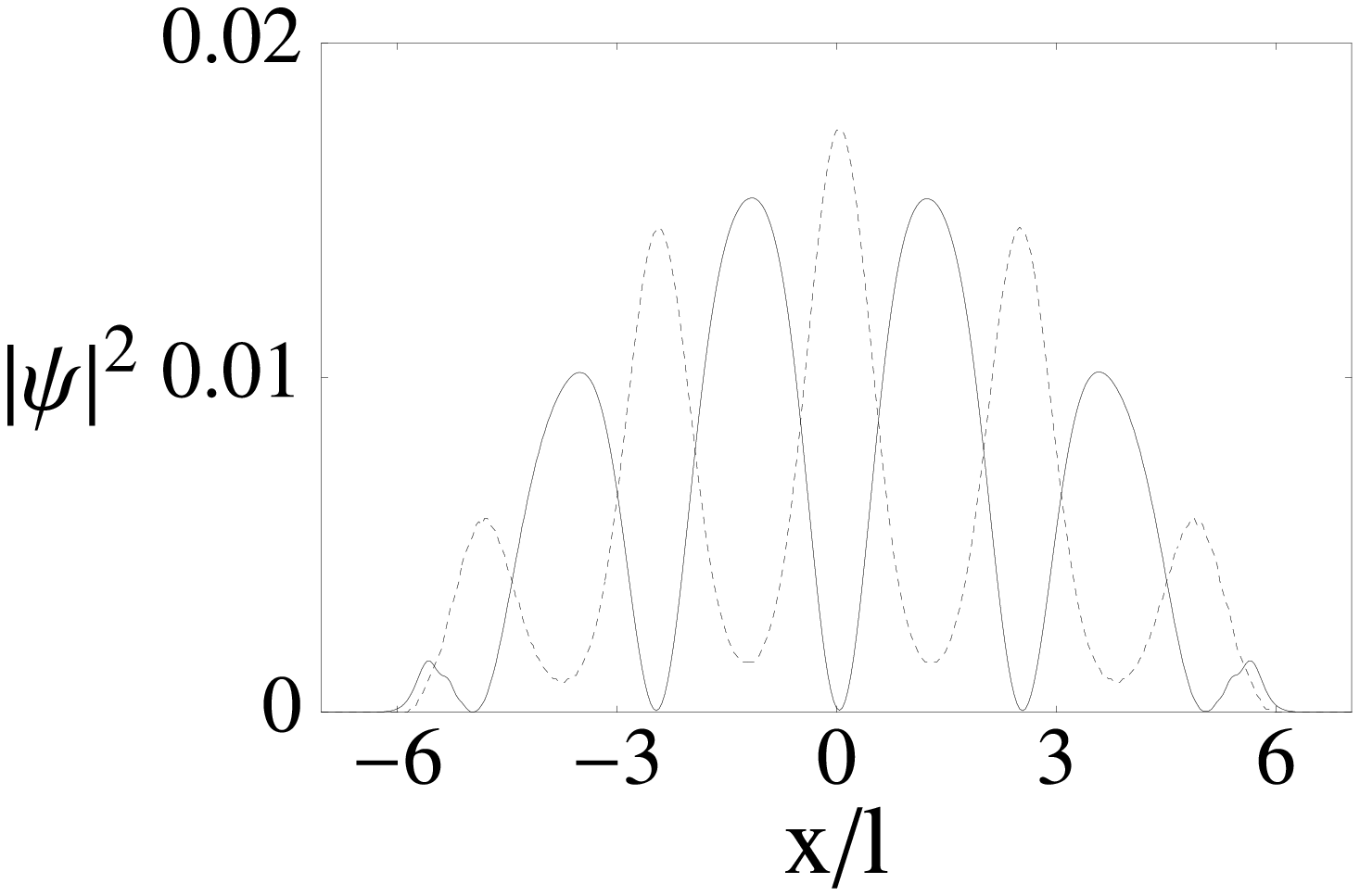}
\end{minipage}
\end{center}
\caption{
The preparation of
rectangular arrays of vortices with the unit circular quantization.
We show the density $|\psi_2(x,y)|^2$ and the phase $\phi_2$ profiles
of level $|2\>$, and the density along the $x$ axis for
atoms in levels $|2\>$ (solid line) and $|1\>$ (dashed
line) for $\lambda=3l$ (left column) and for $\lambda=5l$ (right column).
The circulation around every individual vortex is associated with a 
$2\pi$ change of phase. The flow around the neighboring singularities 
exhibits opposite helicity.
}
\label{fig2}
\end{figure}

Next we consider the preparation of several vortices by means of 
the em fields of Eq.~{(\ref{gpe})}.
In Fig.~\ref{fig2} we show the density and the phase profiles of
rectangular vortex arrays which are obtained with
$\lambda=3l$ (left column) and $\lambda=5l$ (right column).
The plots are evaluated at 
$t=0.025/\omega$, when the em fields are still on. The other 
parameters are the same as in the previous example. 
The topological phases are distributed according to Eq.~{(\ref{topph})}.

\begin{figure}
\begin{center}
\leavevmode
\begin{minipage}{4.2cm}
\epsfig{
width=4.2cm,file=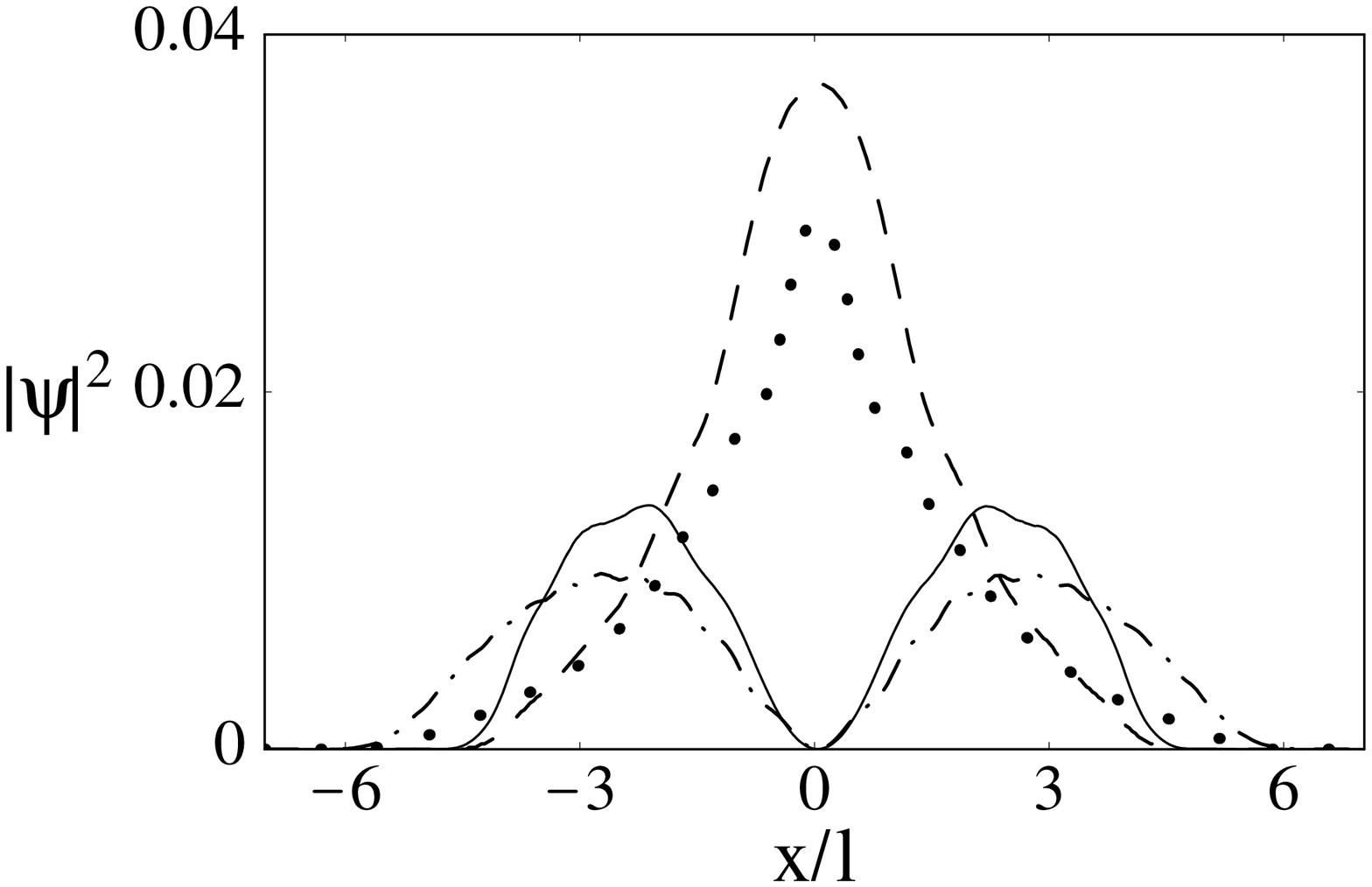}
\end{minipage}
\begin{minipage}{4.2cm}
\epsfig{
width=4.2cm,file=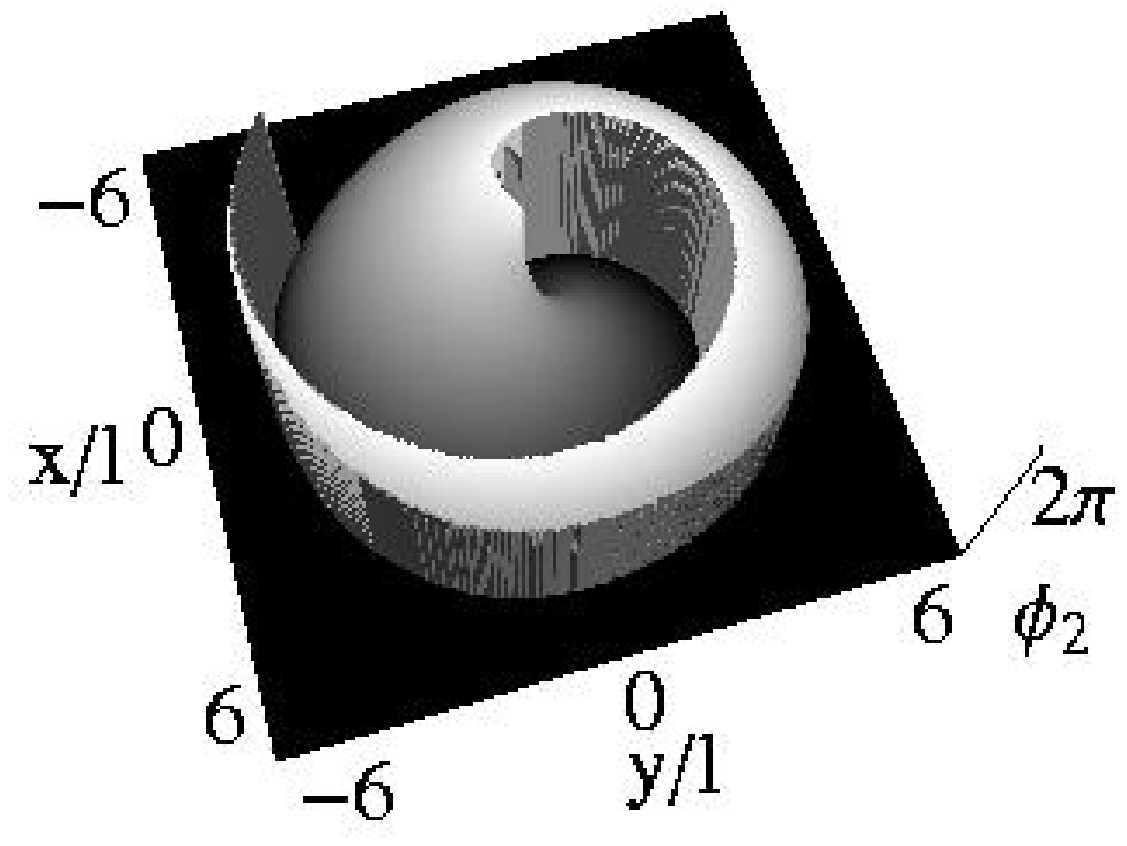}
\end{minipage}
\end{center}
\caption{
The dynamics of a single vortex. We show the phase $\phi_2$ of atoms in level 
$|2\>$ at $t=1.0/\omega$ with the same parameters as in Fig.~\ref{fig1}.
The spiral shape of the phase indicates inward flow due to radial 
oscillations.
The atom density is displayed in levels $|2\>$ (solid line) and
$|1\>$ (dashed line) at $t=1.4/\omega$, when the BEC in $|2\>$ starts
expanding, and in levels $|2\>$ (dashed-dotted line) and $|1\>$ 
(dotted line) at $t=2.5/\omega$.
}
\label{fig3}
\end{figure}

We can investigate the robustness of the prepared vortex states
after the turnoff of the em field coupling. The dynamical stability 
may be studied by integrating GPE with $\Omega=0$ if we ignore the
energetic instabilities due to the interaction of the BECs with thermal
atoms \cite{MAS99,PU99}. In Fig.~\ref{fig3} we show a vortex at 
a later time with the same set of parameters as in the case of 
Fig.~\ref{fig1}. 
The em fields are turned off at $t=0.02/\omega$. At $t=1.0/\omega$
the phase has acquired a spiral shape indicating a radial inward flow.
Although the BECs undergo collective radial oscillations, the vortex
remains well preserved several trap periods. In Ref.~\cite{MAT99} the
radial oscillations were experimentally observed. However, after the
initial shrinking period the vortex expanded and broke up. 
The 2D coherent mean-field picture may not be capable of predicting the
breakup. In the case of a vortex array the dynamical
deformations are more rapid. However, in our studies we have found 
that the characteristic features of the vortex pattern with 
$\lambda=5l$ (Fig.~\ref{fig2}) still exist even at $t=1.0/\omega$.

With em fields we can generate a variety of different topological 
structures. If we are able to use transitions through intermediate levels
involving several photons, we may prepare several vortices in
arbitrary spatial locations with different topological charges.
This construction would then exhibit a periodicity determined by the
wavelength of the em fields. In a more general case the Rabi
frequency from Eq.~{(\ref{ome})} is determined by
\beq
\Omega(\rv)= \bar\Omega_0 \prod_{j=1}^n \{ \sin[\kv_j\cdot(\rv-\rv^{(j)}_0)]
-i\sin[\qv_j\cdot (\rv-\rv^{(j)}_0)]\}^{p_j}\,.
\label{ome2}
\eeq
Here we use $|\kv_j|=|\qv_j|$ and $\kv_j\cdot\qv_j=0$.
The exponent $p_j$ denotes the topological charge of vortex $j$ and $n$
the number of vortex arrays.
For instance, if we use a two-photon transition, we may prepare a
vortex with the topological charge of two. In this case we only have
the $j=1$ term with $p_1=2$ in Eq.~{(\ref{ome2})}.

\begin{figure}
\begin{center}
\leavevmode
\begin{minipage}{4.2cm}
\epsfig{
width=4.2cm,file=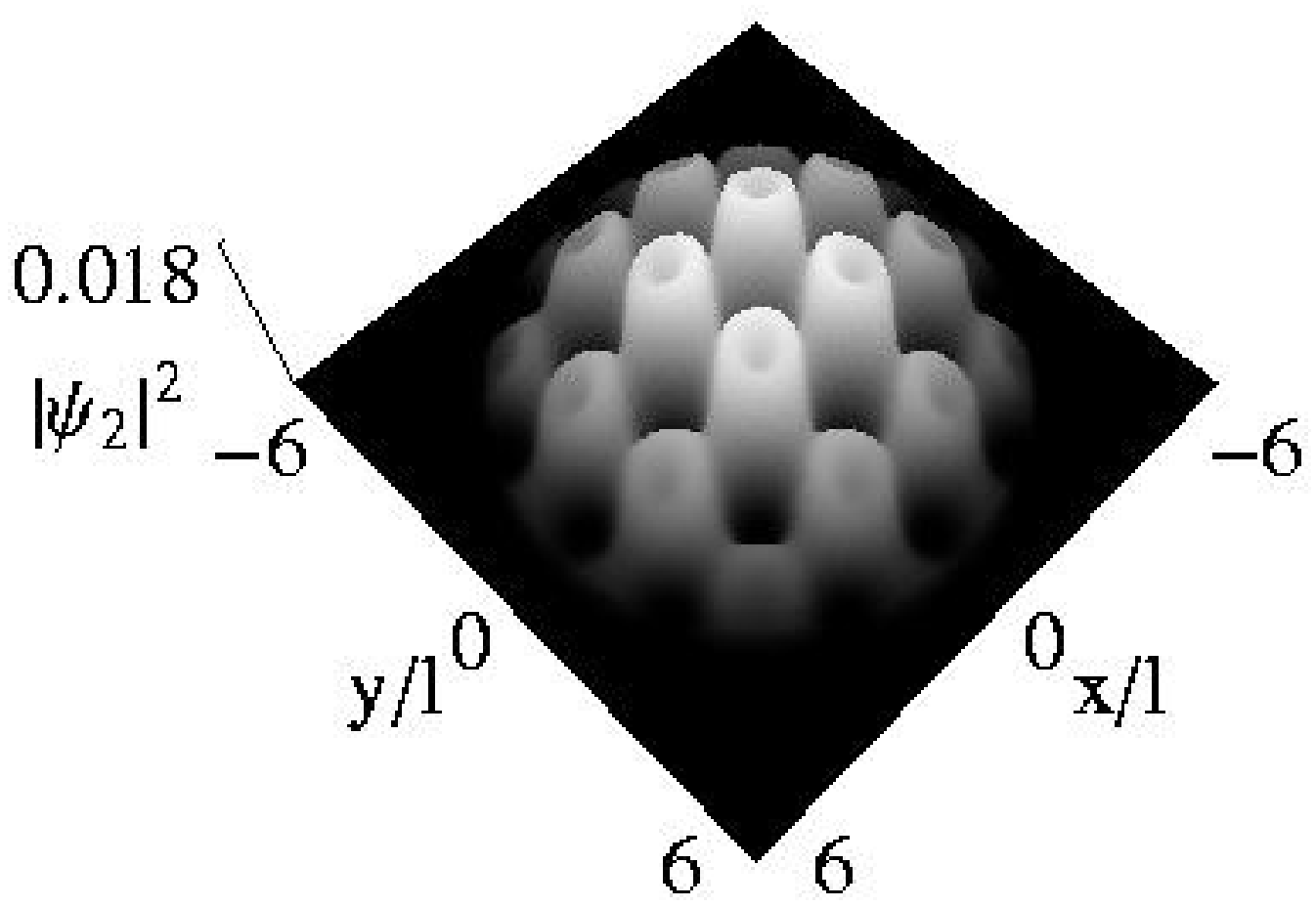}
\end{minipage}
\begin{minipage}{4.2cm}
\epsfig{
width=4.2cm,file=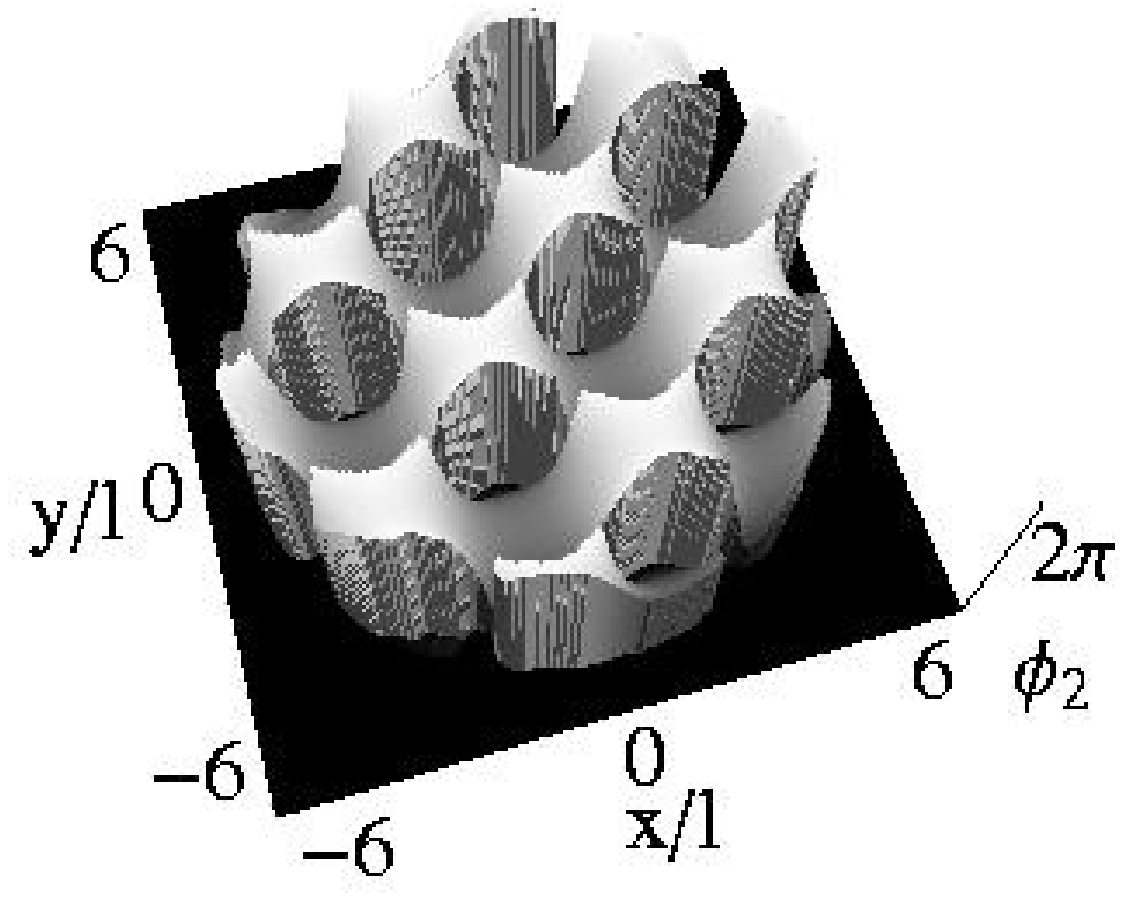}
\end{minipage}
\end{center}
\caption{
The preparation of a rectangular array of vortices with the 
topological charge of two.
We show the density and the phase profiles of level $|2\>$ for 
$\lambda=5l$. The phase profile displays
the $4\pi$ phase windings around the individual vortex lines.
}
\label{fig4}
\end{figure}

In Fig.~\ref{fig4} we show a vortex
array with the topological charge of two at $t=0.02/\omega$, when the
em fields are still on. In this case $\lambda=5l$, $\bar\Omega_0=50
\omega$, and the other parameters are the same as before. 
To reduce the phase noise in the plottings we set the phase in the 
figures equal to zero when the
atom density is very low. Therefore the phase in Fig.~\ref{fig4} is zero 
close to the vortex line. We note the increase or the decrease
in the value of the phase by $4\pi$ as individual vortex singularities 
are encircled.

\begin{figure}
\begin{center}
\leavevmode
\end{center}
\begin{minipage}{4.2cm}
\epsfig{
width=4.2cm,file=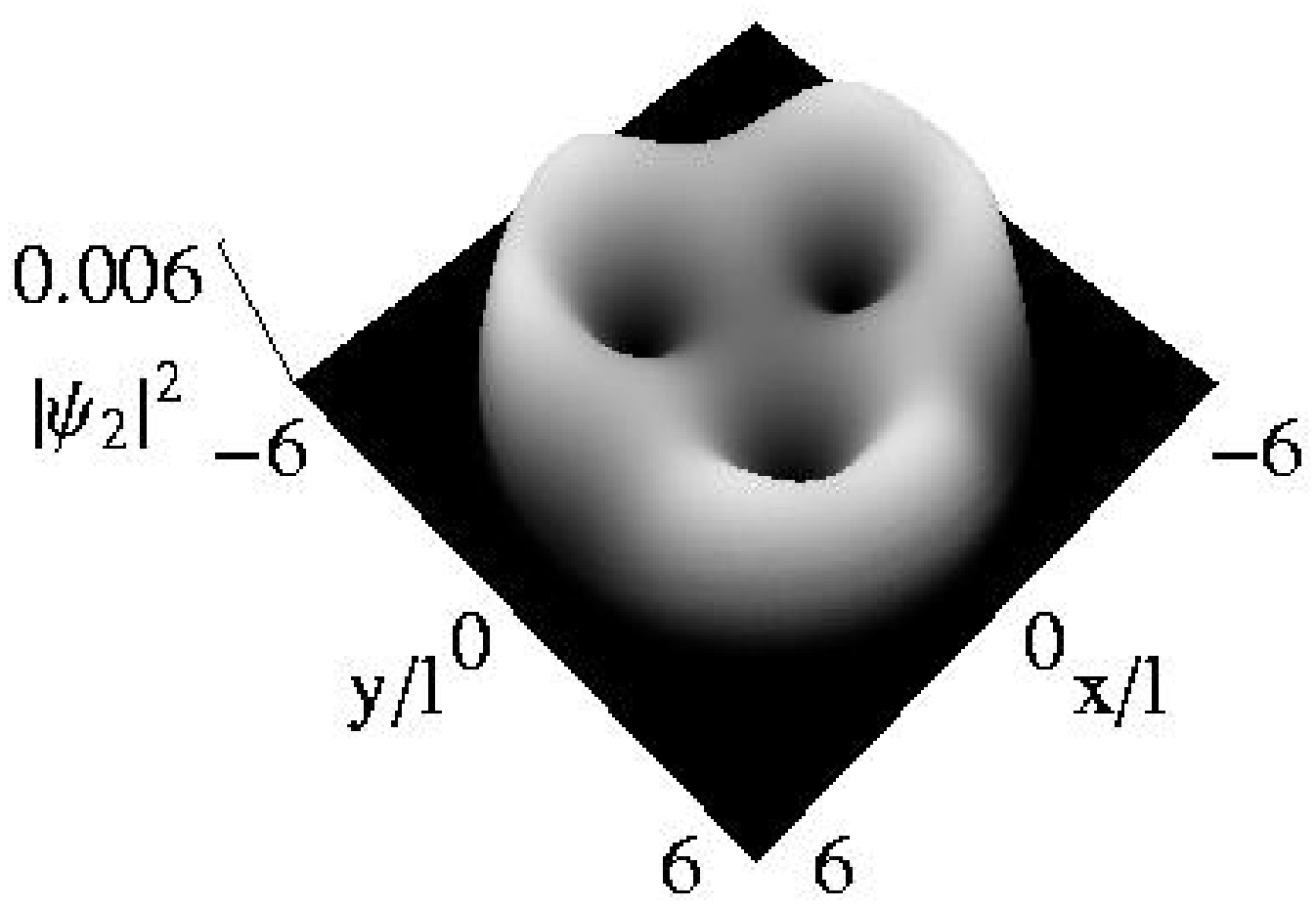}
\end{minipage}
\begin{minipage}{4.2cm}
\epsfig{
width=4.2cm,file=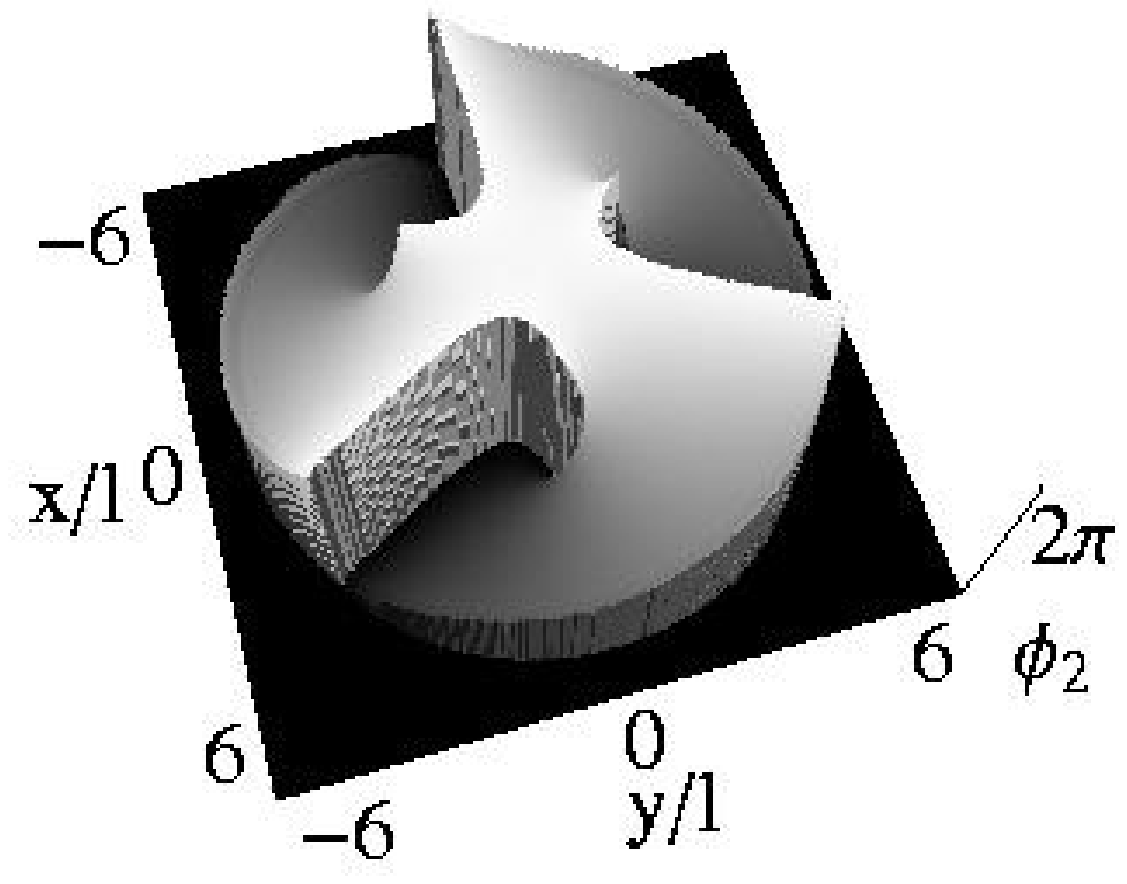}
\end{minipage}
\caption{
The preparation of a three-fold symmetric pattern of vortices with 
the unit circular quantization.
We display the density and the phase of atoms in level $|2\>$.
The circulation around the whole pattern changes the phase by $6\pi$.
}
\label{fig5}
\end{figure}

In Fig.~\ref{fig5} we have prepared a three-fold symmetric pattern of
three vortices with the unit topological charge. This is obtained from
Eq.~{(\ref{ome2})} with $\qv_1/q=\hat{x}$, 
$\qv_2/q=-0.5\hat{x}-0.87\hat{y}$,
$\qv_3/q=-0.5\hat{x}+0.87\hat{y}$, $\rv_0^{(j)}=2.7 l\qv_j/q$,
$\lambda=2\pi/q=20l$, and $\bar\Omega_0=50\omega$. Butts and Rokhsar
\cite{BUT99} have shown that a low-energy state of a rotating BEC can
exhibit a similar symmetry. Nevertheless, the low-energy states cannot
be directly created even in a rotating trap because the rotation  
has to overcome the energy barrier to vortex formation
\cite{FET98,BUT99,FED99}.

\begin{figure}
\begin{center}
\leavevmode
\begin{minipage}{4.2cm}
\epsfig{
width=4.2cm,file=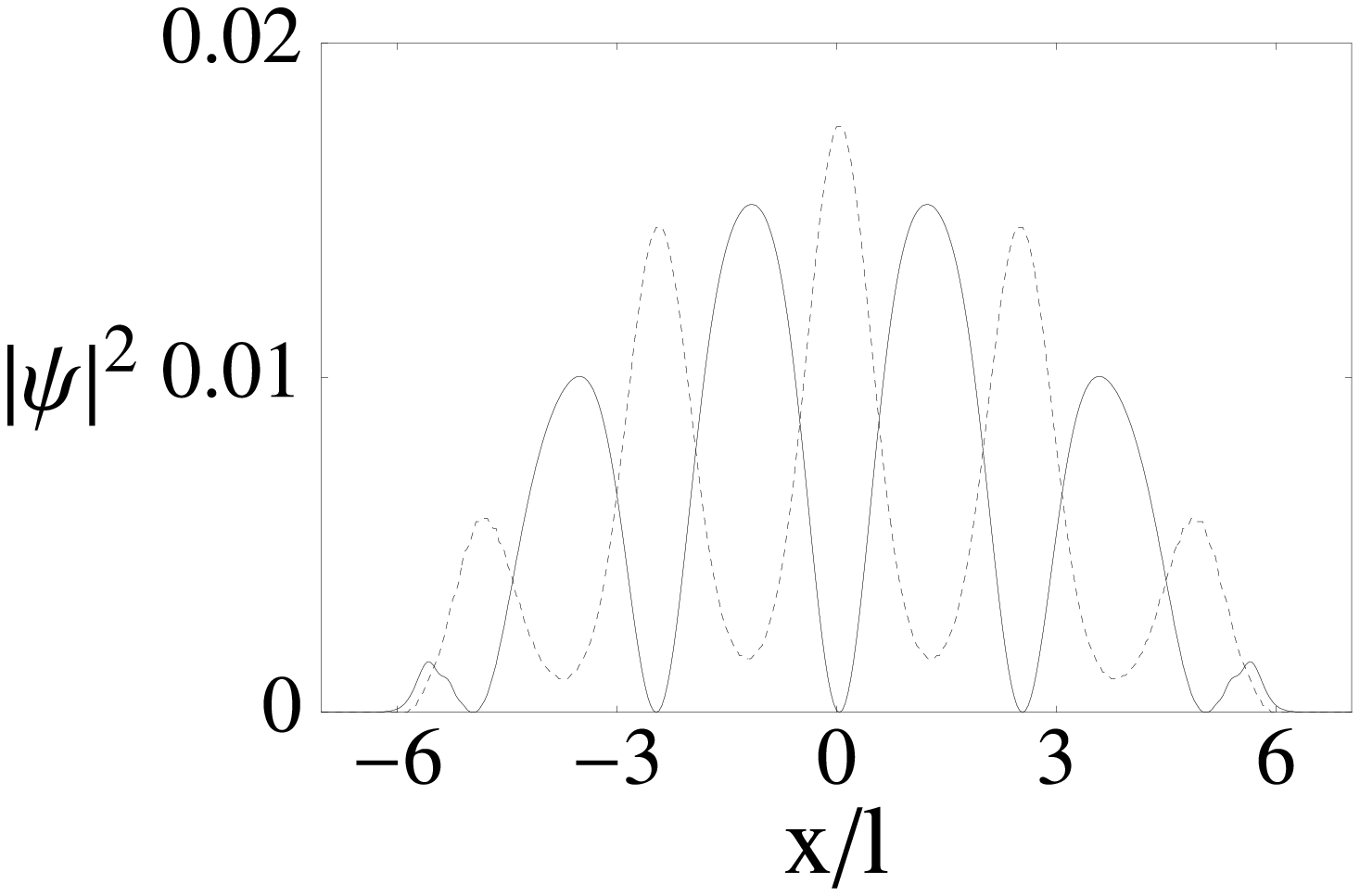}
\end{minipage}
\begin{minipage}{4.2cm}
\epsfig{
width=4.2cm,file=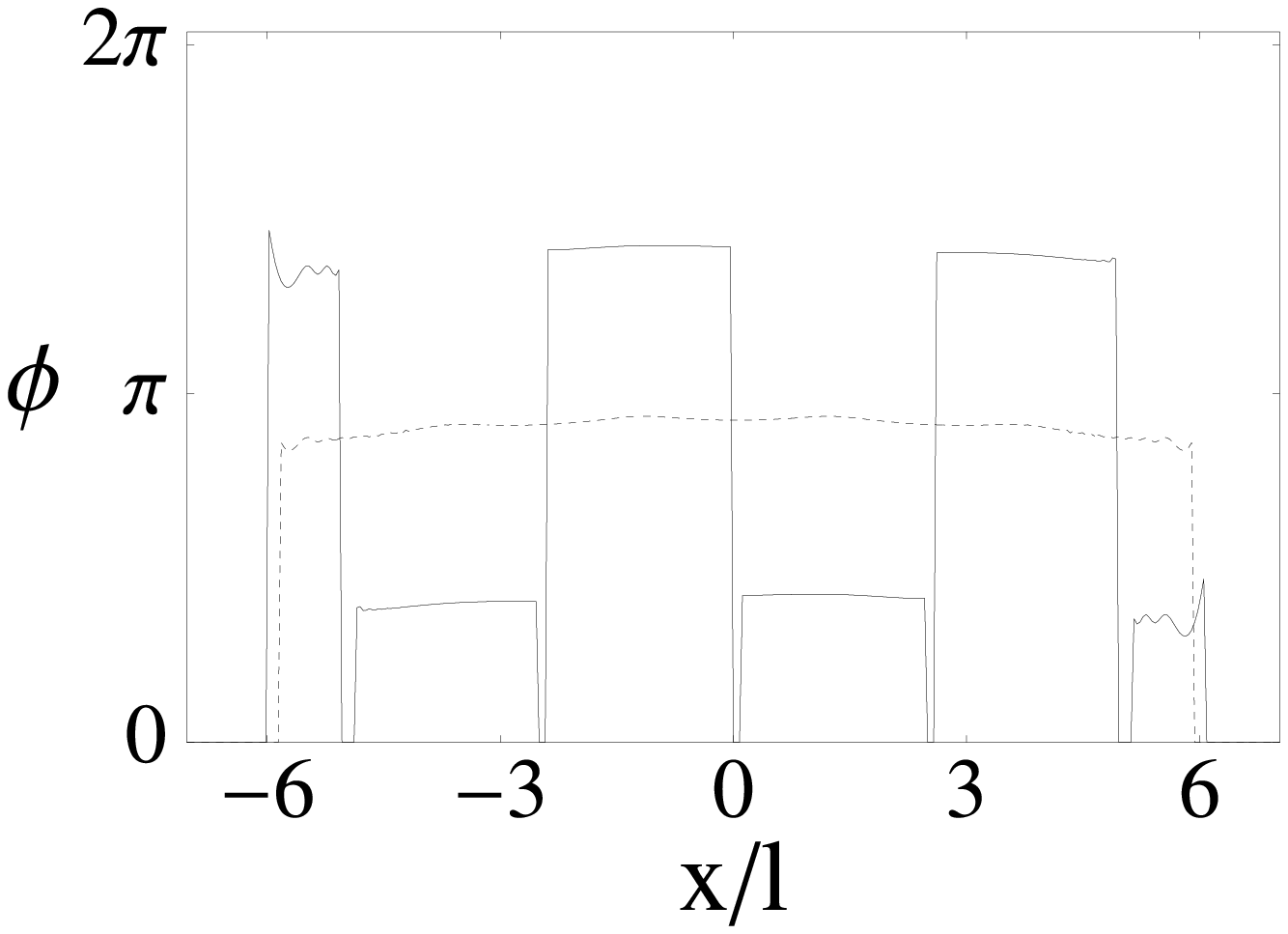}
\end{minipage}
\end{center}
\caption{
The preparation of an array of dark solitary waves.
We show the density and the phase
in levels $|2\>$ (solid line) and $|1\>$ (dashed line)
along the $x$ axis for $\lambda=5l$. The phase graph displays a sharp
phase slip of the order of $\pi$ in the center of the solitary wave
corresponding to the vanishing atom density.
}
\label{fig6}
\end{figure}

As a final example we prepare dark solitary waves \cite{HEL99,BUR99}
with the Rabi frequency $\Omega(\rv)=\Omega_0 \sin(kx)$. In Fig.~\ref{fig6}
we display the atom density and phase at time $t=0.025/\omega$. 
Here we again use the familiar set of parameters, now with $\lambda=5l$ and 
$\Omega_0=50\omega$.

In the previous examples we only considered stationary em fields.
We may also obtain a time-dependent Rabi frequency by generating
a moving standing wave from two nearly counterpropagating em fields
with slightly different frequencies.
In that case we can move skyrmion vortices inside the BECs
along with the moving zero-em-field points in a controlled way. 

We proposed a method of generating periodic vortex arrays
in a binary BEC. The technique relies on a spatially varying standing
em wave configuration which drives atomic transitions between the
BECs. A potential experimental limitation is to find em transitions
with appropriate wavelengths to produce vortex arrays of different
periodicity and number of vortices. Nevertheless, 
the number of practical transition frequencies may possibly be increased 
also by tuning the strength of the trapping potential.

In the present work we have demonstrated only in terms of simple
examples the formation of vortices and solitary waves. More 
complicated em field
configurations could possibly generate more complex topological
structures. The proposed technique could obviously be also generalized
to multi-component BECs.

We acknowledge discussions with J.\ R.\ Anglin. This work was
financially supported by the
EC through the TMR Network ERBFMRXCT96-0066.

\end{document}